# Fabrication Development of a Large Area Grating for Out of Plane Beam Coupling

Jonathan Trisno,[1] Tong Hua Lee,[1] Parvathi Nair S.,[1] You Sin Tan,[1] Ray J. H. Ng,[3] Yingyan Huang,[4] Seng Tiong Ho,[4,5] and Joel K. W. Yang[1,2,*]

[1] *Singapore University of Technology and Design, Singapore, 487372, Singapore*
[2] *Institute of Materials Research and Engineering (IMRE), A*STAR, 138634, Singapore*
[3] *Institute of High Performance Computing, A*STAR, 138632, Singapore*
[4] *OptoNet Inc., Evanston, IL, 60201, USA*
[5] *Northwestern University, Evanston, IL, 60208, USA*
\* *joel_yang@sutd.edu.sg*

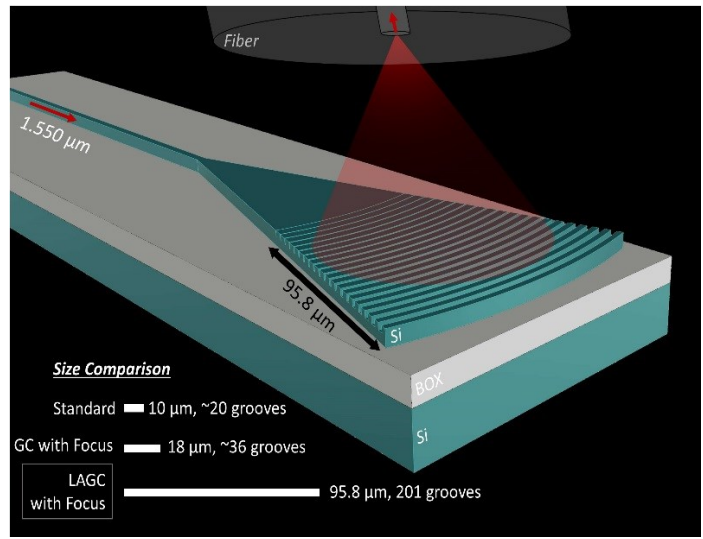

**Abstract:** We develop a single-layer waveguide surface grating structure to vertically couple near infrared (NIR) light at ~1.55 µm wavelength from a large area (~100 µm length scale) Si waveguide on a Silicon-On-Insulator (SOI) substrates to free-space for high-power laser applications. Our design approach is based on the optimization of local emission angles and the out-coupling intensities. Simulation results show that a focal spot with a $1/e^2$ width of 3.82 μm can be achieved at the desired focal position, with 33% (-4.81 dB) simulated source to free-space focusing efficiency, while initial measurements show an efficiency of 22% (-6.58 dB).

## 1.Introduction

High power diode and fiber lasers have been realized via beam combining of thousands of elements with output powers in the range of kW [1,2]. Compact lasers with high-power output could potentially be achieved using hybrid silicon lasers, *i.e.* silicon and III-V materials with output power per gain element in the range of tenths to hundreds of mW [3,4]. Using beam combining techniques, output from 300 to 600 gain elements (with per-gain element output of up to ~1 W) could be combined to realize a high-power semiconductor laser with hundreds of Watts output power. One challenge of a high-power laser is the size of its output channel would require larger gratings for out-of-plane coupling.

In this work, as an exemplary demonstration we investigate the design and fabrication of a large area grating coupler (LAGC) with a dimension of ~100 μm that takes input from (or gives output to) a 1 μm-wide connecting waveguide that is tapered out to a dimension of ~61.4 μm at the first grating line, and the 1 μm width is chosen only for convenience of simulation and experimental data collection. The main focus is the design and performance of the ~100 μm large grating that would be able to connect optical beam between an optical fiber and a wide waveguide up to 100 μm in width. While reports have shown that single-layer waveguide surface grating elements can provide both out of plane coupling and focusing without the need of a lens [5–8], these designs were limited to grating dimensions in the order of 10-20 microns. Scaling the grating dimensions by an order of magnitude larger (from 10's microns to 100's microns in size) creates new challenges in terms of grating design, electron beam lithography (EBL), and fabrication processes that have to be carefully strategized and optimized in order to achieve the desired beam interference and nanofabrication fidelity required to obtain the intended beam's upwards output coupling and focusing.

Large photonic structures (hundreds of μm to mm scale) are typically fabricated using photolithography techniques that allow for relatively fast fabrication, but with limited resolution [9–11]. For example, a large-scale nano-photonics phased array (NPA) structure of 64 x 64 element has been fabricated with a 193 nm oil immersion lithography [11]. Though the device dimension is large (~500 μm length scale), the per-element size is ~3 μm and contains only 5 grooves. Conversely, a large focusing grating requires hundreds of grooves with varying feature sizes, and requires a resolution down to ~10 nm scale. This challenge is suitably tackled by EBL due to its high resolution [12–15]. Here, we address the fabrication challenges in controlling the dimensions of structures accurately across ~100 μm length scale, which is much larger than the typical size of a grating coupler (GC). Our fabrication strategy allows high resolution gratings with large variations in feature size to be patterned and etched.

## 2. Results and Discussions

### 2.1 Design of Large Area Apodized Grating

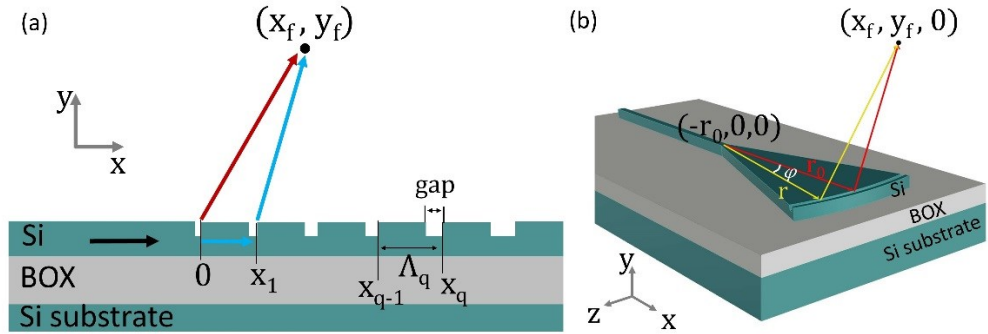

Fig. 1. Schematic of grating design to achieve focusing: phase-matching conditions (a) in XY plane by periodicity modulation, (b) in XZ plane by curving the grating with variable radius r(φ).

The grating consists of a waveguide on SOI with Si device thickness= 500 nm and Buried Oxide (BOX) thickness= 3 μm. The input channel is a ridge waveguide with a 1 μm width which tapers out into a 61.4 μm width at the first grating line. The waveguide is fully etched on the sides, while the grating grooves are partially etched with a depth of 100 nm. We define periodicity ($\Lambda_q = x_q - x_{q-1}$) as the distance from the edge of one groove to the corresponding edge of the previous groove, where q is the groove number, and the duty cycle $\rho_q$ is the ratio of the gap to the periodicity of groove q. A 2-D focusing effect is achieved by varying the periodicity. Figure 1a shows the schematic highlighting the phase matching condition used [16]. The guided mode in the Si waveguide propagates along the x axis and is scattered into the y axis by the grooves. Each groove is positioned such that the scattered light rays constructively interfere at a desired focal point ($x_f$, $y_f$) as determined in Eq.1, where q is the groove number (0,1,2,…,200), $x_q$ is the position of groove q relative to the first groove ($x_0$= 0), k is the wavevector of light at vacuum, $\rho_q$ is the duty cycle of the groove q, $n_{air}$ is the refractive index of air ($n_{air}$= 1), $n_{Si}$ is the index of un-etched region of silicon ($n_{Si}$= 3.269), $n_{gap}$ is the index of etched region of silicon ($n_{gap}$= 3.186), and $n_{grating}$ is the effective refractive index of the grating which is influenced by $\rho$. Focusing in the third dimension can be achieved by curving the grating (Eq.2), as shown in Fig. 1b.

$$kn_{air}\left[x_f^2 + y_f^2\right]^{1/2} + q2\pi = kn_{grating}x_q + kn_{air}\left[(x_f - x_q)^2 + y_f^2\right]^{1/2} \quad (Eq.1)$$

$$\text{where } n_{grating} = (1-\rho_q)n_{Si} + \rho_q n_{gap}$$

$$kn_{Si}r_0 + kn_{air}\left[(x_f - r_0)^2 + y_f^2\right]^{1/2} = kn_{Si}r + kn_{air}\left[(x_f - r\cos\varphi)^2 + (r\sin\varphi)^2 + y_f^2\right]^{1/2}$$
(Eq. 2)

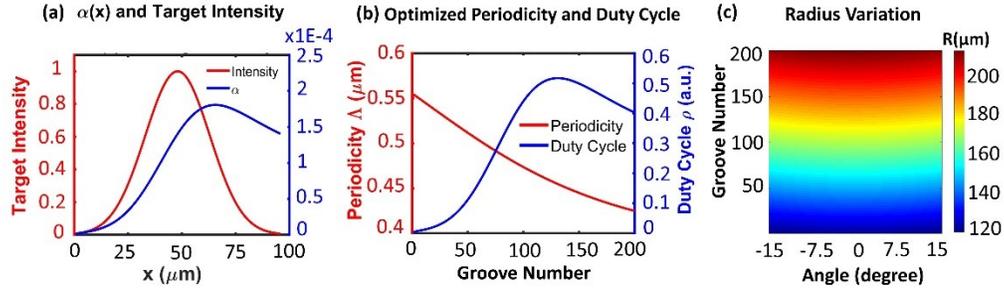

Fig. 2. Calculated design parameters: (a) Target intensity profile (red) is used to determine the required scattering strength α (blue); (b) Optimized periodicity (red) and duty cycle (blue); (c) Variation in radius of curvature for different groove number.

Next, the duty cycle ρ(x) is calculated to achieve the target intensity profile above the grating. Controlling this intensity profile enables a strong overlap between the focal point intensity profile and the fiber mode, enhancing fiber coupling efficiency [5]. P(x) in Eq. 3 describes the power of light that propagates in the waveguide and the grating region. As light propagates through the grating, it will be continuously diffracted, thus leading to an attenuation of P(x). The attenuation profile of P(x) can be fitted using an exponentially decaying function, shown in Eq. 3, with α described as the grating groove scattering strength. I(x) is the intensity profile of the diffracted light, as derived in Eq. 4. I(x) is mapped to a Gaussian profile (red line Fig. 2a) corresponding to the far-field projection of the target circular Gaussian beam with diffraction limited width of 2.7 um at $1/e^2$ point of the intensity under a Numerical Aperture of N.A. = 0.35 (the $1/e^2$ beam width is 2.7 μm). At the surface of the grating, this profile (I(x)) corresponds to a beam-width of 61.3 μm at $1/e^2$ point of the intensity or ~96 μm full beam-width, which basically gives the required physical size for the surface grating along the propagation direction. The locally varying values of ρ(x) are generated based on the needed scattering strength α (blue line in Fig 2a) extracted from Eq.6 with C value adjusted based on groove etching depth, in our case C=2.77x$10^3$.

The final design parameters of the grating are depicted in the plots of the modulated grating periodicity and duty cycle in Fig. 2b (red line and blue line respectively). The result is based on focusing at a target spot ($x_f$= 30 μm, $y_f$= 80 μm), with emission angle of -10 degree. We use a slight backward emission angle as it will give better efficiency than positive angle, while also cut off higher order modes [5,17,18]. Perfectly vertical emission angle (0 degree) will cause a drop in efficiency due to strong back-reflections, while a solution for this usually requires multi-layer gratings [19–21]. The periodicity is non-linearly decaying while the duty cycle shows a similar shape to the scattering strength. The duty cycle profile indicates that the front section of the grating will have a very small gap, which will increase and reach its maximum at grating groove number 131 with ρ=0.52 and will then decrease towards the end section of the grating. Figure 2c shows the radius of curvature as a function of angle for different groove numbers. For a single groove number, the radius varies up to ~2.5 μm with the 0 degree element having longer radius compared to other angles.

$$P(x) = P_0 \exp(-2\alpha x) \qquad (Eq. 3)$$

$$I(x) = -\frac{dP(x)}{dx} = P(x)\left(2\alpha + 2\frac{d\alpha}{dx}x\right) \qquad (Eq.4)$$

$$\left(\alpha + \frac{d\alpha}{dx}x\right) = \frac{1}{2}\frac{I(x)}{P_0 - \int_0^x I(x')dx'} \qquad (Eq.5)$$

$$\rho(x) = C\sin^{-1}(a(x)) \qquad (Eq.6)$$

*2.2 Large Area Grating Fabrication*

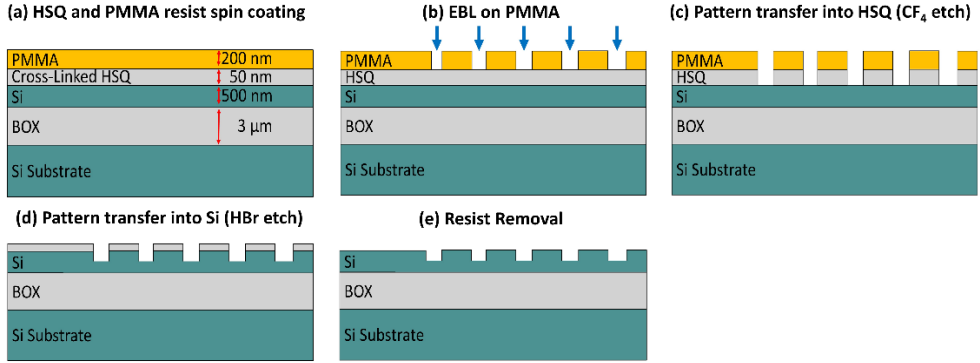

Fig. 3. Schematics of the fabrication process flow: (a) Spin coating of resists; (b) EBL on PMMA; (c) CF$_4$ etch into HSQ; (d) HBr etch into Si; (e) Resist removal.

The fabrication of the grating device comprised two aligned electron beam lithography (EBL) and etching steps. We used a double-layer mask, consisting of polymethyl methacrylate (PMMA) and silicon oxide resists. A positive resist was chosen for the lithographic process, as it allows the layout to be patterned most effectively because less area will need to be exposed. PMMA was used due to its high resolution limit and relatively high sensitivity [22], thus reducing patterning time when compared to negative resists like hydrogen silsesquioxane (HSQ). PMMA, however, is not a good mask to etch into Si due to its low selectivity. A cross-linked HSQ was used as an oxide mask to etch into Si. As the grating section requires a partial etch while the waveguide section requires a full etch, a double-etch procedure was used. The grating, along with alignment markers was made by pattering using EBL and etching into Si. The waveguide was then patterned during a second EBL step via a three-point alignment process, before being fully etched. We observed misalignment errors to be <50 nm, which is relatively small compared to the period of the grating and the operating wavelength, thus this error will have negligible effects on grating performance.

The EBL process was optimized by solving several lithographic errors that occur in patterning over a large area, such as stitching issues [23], line-edge roughness, kinks, dose non-uniformity, and long patterning time. Careful tailoring of pattern layout and dose factor is required for fabricating an apodized grating with varying periodicity and duty cycle as the accuracy of the fabricated grating coupler is strongly related to the properties of EBL used during exposure [23,24].

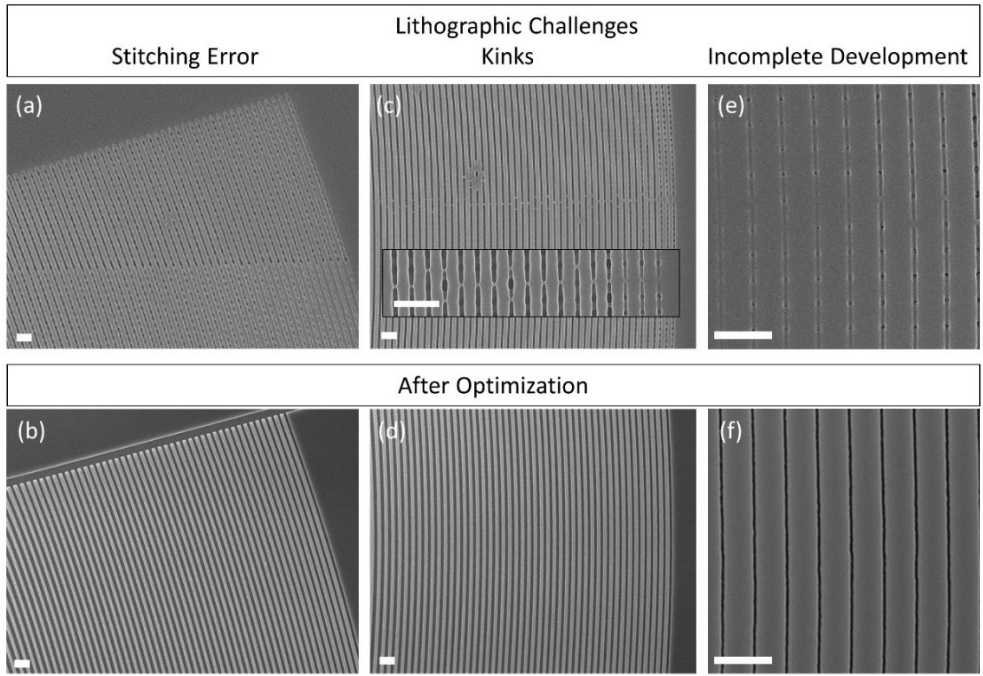

Fig. 4. Scanning Electron Micrograph (SEM) of PMMA mask highlighting lithographic errors and optimized structure: (a, b) stitching errors, (c, d) kinks and systematic errors due to e-beam line exposure, (e, f) incomplete development requiring layout and dose factor (DF) adjustment. (scalebar= 1 μm)

When patterning over a large area, a pattern is normally divided into smaller write fields (WFs), which are then combined by stitching through stage movement, potentially introducing stitching errors. To maximize the patterning resolution, we first used a 100 μm x 100 μm WF, which required the grating to be divided into 2 WFs. Despite the stage being equipped with a laser interferometer, stitching errors can be observed at the WFs boundary (Fig. 4a). To completely remove stitching issues in the grating region, we increased our WF size to 200 μm x 200 μm (Fig. 4b) to fit the entire grating. We observed no significant resolution loss, as we were able to fabricate single pixel line (SPL) with linewidth < 20 nm.

Figure 4c shows systematic errors from line exposure to form the curved grating. The error was the result of how the EBL software interpreted and scanned the curved-grating layout.

The EBL scanning and exposure of the grating structure on PMMA is based on the curved-grating layout which is generated via line-connected discrete points (previously generated using Eq. 2). During scanning, the dose will build-up when the beam slightly changes its direction at these points, causing the systematic over-dose error. The error is more visible on grooves with larger gaps as the dose build-up is more significant. This becomes increasingly obvious in a sensitive resist like PMMA. The effect was minimized by pre-baking the PMMA to reduce its sensitivity, as pre-baking removed the solvents. With reduced sensitivity, the resist was able to handle higher exposure doses (by almost 1.5x), allowing us to introduce boundary lines with a higher dose factor and to slightly reduce the dose factor of the fillings to minimize the error and obtain well-defined boundaries (Fig. 4d) with no over-dose structure observed.

The apodized design of our grating duty cycle corresponds to a very large gap size variation (from 235 nm down to 5 nm), which requires the fabrication layout and dose factor (DF) to be carefully tailored for accurate patterning and to compensate for proximity effect. To achieve this, we divided the patterning layout into different sections. The grooves on the first section (groove 0 to 25) were patterned using 1 single pixel line with the gap size varied using the DF, creating the smallest gap size of 16 nm, slightly larger than the required smallest gap size of 5 nm due to lithographic resolution limit. The grooves in the second section (groove 26 to 79) were patterned as a series of single pixel lines with a line spacing of 5 nm, that allowed a precise control of gap size by tuning the number of lines and the DF. This method, however, was very time consuming, as the beam was blanked between the end of one line and at the start of a new line. The beam was also 'dynamically compensated' at the start of each new line by positioning the beam slightly further from the original position to allow the beam to ramp up to its normal speed before starting to expose the line (settling time), increasing patterning time.

As pattern optimization requires many design tests and dose optimization steps, a long patterning time for a single mask prevents the design to converge effectively. To reduce patterning time, grooves in the last section (groove 80 to 200) were patterned using filled polygons with 2 SPL lines as boundaries. This method allowed the gap to be drawn with lines in a continuous meander. At the end of the patterning of a line, the beam moved to the new line position and continued to pattern in the opposite direction without being blanked. This resulted in a faster patterning as settling time was minimized. The 2 SPL lines boundaries were introduced with larger DF to produce a well-defined edge with minimum roughness. The lower DF of the filling minimized the E-beam exposure proximity error, allowing accurate control over the gap size, as well as removing the systematic error previously mentioned. Figure 5 shows the schematics of the patterning methods for different sections. By introducing the last patterning method, the grating patterning time was significantly reduced by 75.6%, from 45 minutes down to 11 minutes.

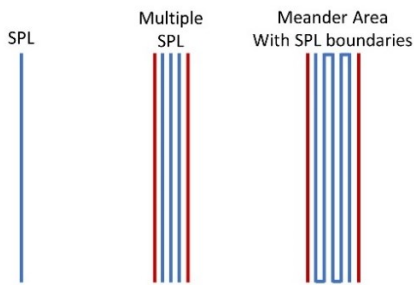

Fig. 5. Schematics of patterning method of different grating sections.

The etching process for HSQ and Si was performed using an Inductive Coupled Plasma-Reactive Ion Etcher (ICP-RIE). The pattern was transferred from the PMMA mask into cross-linked HSQ via $CF_4$ dry-etch, with 45 sccm gas flow. The patterned HSQ was then used as the hard-mask for silicon etching using hydrogen bromide (HBr) chemistry because of its high etching selectivity for Si etch with Si oxide as a mask [25,26]. The detailed etch conditions can be found in the Method section.

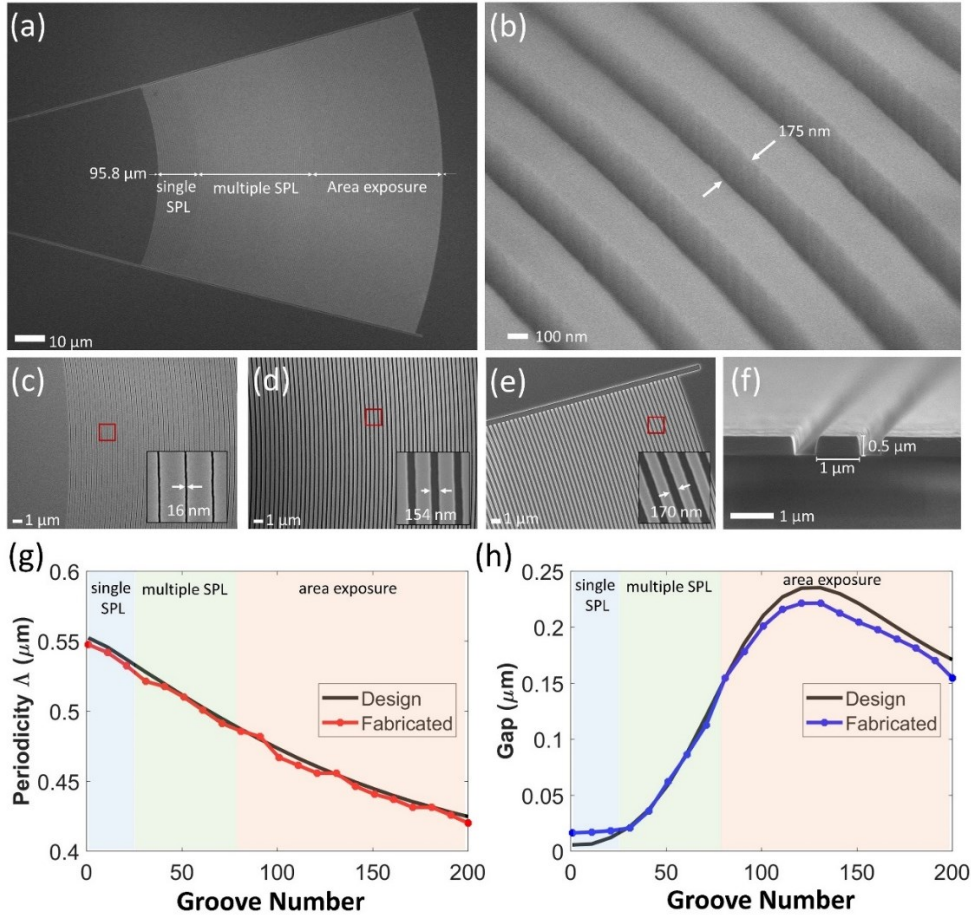

Fig. 6. SEM of Si LAGC on SOI showing: (a) the whole apodized grating with varying periodicity (Λ) and duty cycle (ρ), (b) tilted view of the grating, (c) front-section patterned with single pixel lines (SPL), inset showing 16 nm linewidth, (d) mid-section patterned with proximity corrected multiple SPLs, (e) end-section patterned with proximity corrected area exposure to reduce patterning time, (f) cross-section of waveguide. The patterning accuracy is shown in (g) for periodicity and (h) for gap size.

Figure 6a shows the fabricated LAGC on SOI, with its multiple patterning sections. The total grating length is 95.8 μm, with 13.4 μm patterned with single SPL, 38.9 μm with multiple SPL, and 43.5 μm with area exposure. Figure 6b shows the grooves etched into the Si layer. Figure 6(c-e) shows the sections patterned with the single SPL, multiple SPLs, and area exposure methods respectively. Figure 6f shows the cross-section of the input waveguide.

To determine the patterning accuracy of the optimized design with tailored layout and DF, we took SEM micrographs of the grating at 21 different locations with 50k X magnification and measured the periodicities and gap sizes. Figure 6(g, h) shows the designed and fabricated periodicity and gap respectively. The periodicity of the fabricated grating shows a close match with the design with a maximum error of 7 nm and a standard deviation of 2 nm. The gap shows a maximum error of 17.3 nm with a standard deviation of 5.2 nm. This error is marginal as it is only 8% of the corresponding gap (212.5 nm). Upon closer inspection, the multiple SPL region demonstrated a very good fit between the designed and fabricated structures, which indicated a good accuracy of this patterning method. The error mostly occurred at the single SPL region, as well as the area exposure region. The error at the single SPL region was caused by the lithographic resolution limit of our EBL, so the smallest line-width achieved was 16 nm, which corresponds to ~10 nm error in gap size. The maximum error of 17.3 nm, however, occurred in the area exposure region, indicating that there is a trade-off between patterning time and patterning accuracy.

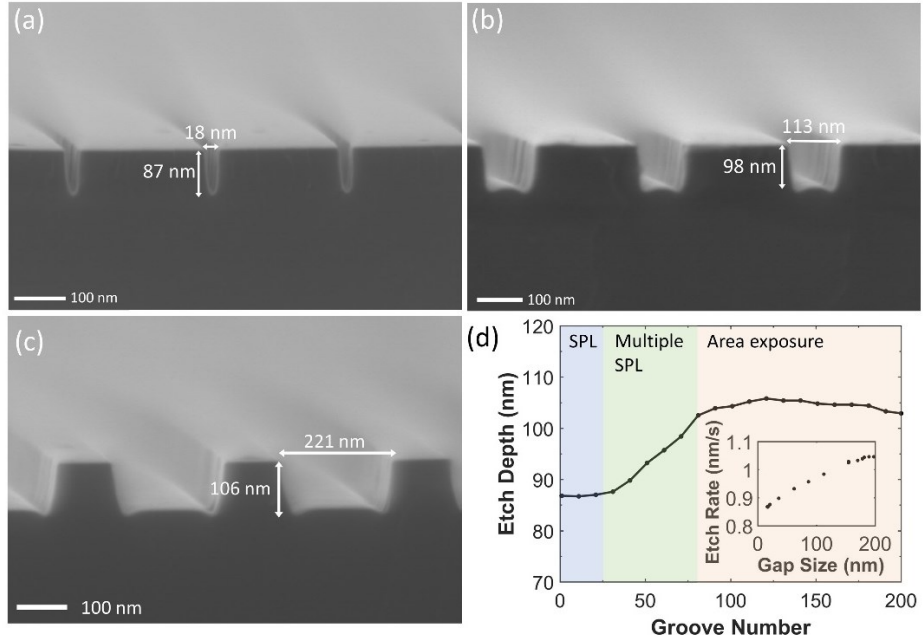

Fig. 7. Cross-section SEM of (a) SPL section, (b) multiple SPL section, and (c) area exposure section. Etch depth analysis is shown in (d), with inset showing etch rate.

Reactive ion etching (RIE) of trench structures with varying gap sizes has been known to exhibit RIE-lag effect, where smaller gaps will have a slower etch rate, and hence will be etched to a shallower depth when subject to the same etch duration [27,28]. Variation in the etch depth causes the effective refractive index of each groove to deviate from the intended index, impacting the device performance in terms of their focusing accuracy and efficiency. Figure 7 shows the etch depth analysis of our gratings. Figure 7(a, b, c) shows the cross-section SEM images of the single SPL section, multiple SPL section, and area exposure section respectively, with measured etch depths of 87 nm, 98 nm and 106 nm (corresponding to groove no 20, 70, and 120). It can be observed that the etch depth increases as the gap size increases. Microtrenching effect can also be observed in larger gaps (Fig 7b, c) due to faster etch rate near the sidewall. The effect can be attributed to the reflection of the ions from the slanted sidewall, that was caused by post-developed resist profile or faceting effect of the etching mask [29–34].

Figure 7d shows measurements on 21 different grooves with an increment of 10 in the groove number (groove 0, 10, 20, …, 180, 200). The depths were measured at 5 different locations for each groove and averaged, exhibiting standard deviation ranging between 0.44 nm and 1.6 nm. The single pixel section (groove 0-25), which has gap sizes around 16 to 20 nm shows shallower etch depths of 86 to 87 nm. The shallower etch depths as compared to the designed etch depth of 100 nm, provide slight compensation for the gap error due to fabrication limitation as they compensate the deviation of effective refractive indices. The multiple SPL region (groove 26-79), which gap sizes vary significantly from 20 to 154 nm, shows etch depths that increase from 87 to 103 nm. The area exposure section (groove 80 to 200), which has gap sizes around 150 to 221 nm, shows etch depths around 102 to 106 nm. The deeper etch depths will slightly compensate for the gap fabrication error as shown previously in Fig. 6h. The significantly lower variation in the etch depth as compared to the multiple SPL section, shows that grooves with gap size above 150 nm are less impacted by the lag effect. The inset in Fig. 7d shows the etch rate for different gap sizes. The variation in etch rate is more pronounced for smaller gap sizes, while gaps larger than 150 are much less impacted by the RIE-lag effect. The smaller gaps etch slower as it is more difficult for the ions/reactive elements to pass through the gaps, while the etch byproducts also cannot diffuse out as easily.

*2.3 Grating Performance*

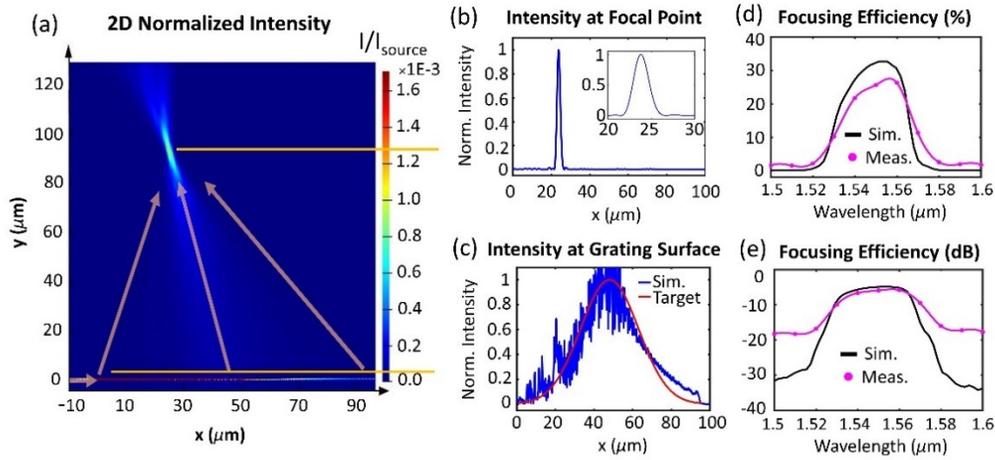

Fig. 8. 2-D FDTD simulation for λ=1.550 μm: Normalized intensity profile **(a)** showing focusing effect at focal point (23.8, 92.2) μm, **(b)** at focal plane y = 92.2 μm, showing $1/e^2$ width of 3.82 μm, **(c)** at grating surface showing match between simulation result and target profile. The simulated (black) and measured (magenta) focusing efficiency as a function of wavelength is shown in figure **(d)** and **(e)** in percent and dB respectively.

The LAGC performance is assessed by evaluating its focusing accuracy and efficiency. When modeled based on the ideal design, the 2-D FDTD simulation shows that the grating focal point is located at $x_f$= 27.5 μm and $y_f$= 79.1 μm, a slight shift from the design target ($x_f$= 30 μm, $y_f$= 80 μm) with 38% focusing efficiency (-4.20 dB) for λ= 1.550 μm. Focusing efficiency is defined as the power measured at the focal point divided by the incident light at the grating start.

However, when the error in gap size and variation in etch depth are considered (modeled closer to fabricated device), the focusing accuracy and efficiency are affected. Figure 8a shows the focal point located at $x_f$= 23.8 μm and $y_f$= 92.2 μm, with the Numerical Aperture (NA) corresponding to ~0.32. The emission intensity at 0.8 μm from the grating surface (Fig. 8c) shows a Gaussian profile (blue) compared to target (red). Figure 8b shows the intensity profile at the focal plane y=92.2 μm, with $1/e^2$ width= 3.82 μm at the focal point, compared to the target Gaussian beam width of 2.7 μm. The peak simulated focusing efficiency (Fig. 8d, e in black) from Si waveguide to free-space is 33% (-4.81 dB) for λ= 1.554 μm, with a 3 dB bandwidth of 34 nm centered at λ= 1.552 μm. The loss analysis shows that 46% of the light is diffracted downwards into the oxide substrate, 9% is diffracted to higher order, 5% is diffracted back towards the source, and 6% is transmitted through the grating. The efficiency can be significantly optimized by introducing an embedded reflector (e.g. metallic mirror or Bragg reflector) or by tuning oxide thickness to recover downwards diffracted light.

Measured efficiency of the device is shown in Fig. 8d,e in magenta. An input light (λ= 1.5 μm to 1.6 μm with 0.01 μm interval) was generated using a tunable laser source (Santec TSL-550) with a lensed fiber (SMF28) output. The power from the fiber was measured to be 15.8 mW. The light from lensed fiber (spot size ~2 μm) was then aligned and edge-coupled into our Si waveguide with efficiency ~10% (-10 dB), taking into account insertion and waveguide loss, resulting in ~1.58 mW incident at the grating start. The edge coupling efficiency was estimated via measurements of straight waveguides with different lengths. The propagating light was then diffracted by the LAGC. The diffracted light power was measured using a power meter (Thorlabs PM122D) behind a 100 μm aperture. The power meter was mounted on a 3-axis micrometer stage, vertically placed as close as possible such that the aperture was at ~150 μm from the sample surface, at which height the spot size would have enlarged to ~50 μm. Hence a 100 μm is suitable to block most random scattering while allowing the focused beam to be captured. The power meter was translated in the horizontal plane until maximum reading was observed for $\lambda$= 1.55 μm. The measurement results show a maximum focusing efficiency of 25.3% (-5.97 dB) for $\lambda$=1.56 μm, with a 3 dB bandwidth of 36 nm centered at $\lambda$= 1.556 μm. Our calculations show that scattered light would account for only a ~3% drop of efficiency of the light collected within the aperture. We thus estimate the fabricated device free-space focusing efficiency to be 22% (-6.58 dB). The discrepancies between the simulation and measurement results are mainly caused by focusing loss in the third dimension (as the simulation is only 2-D).

The simulation results show that the design method allows for relatively precise focusing at the target focal spot and for intensity control of the beam. The grating periodicity and radius of curvature can be modified to accommodate the NA of the fiber, while the duty cycle can be adjusted to produce a far-field intensity profile that overlaps the fiber mode profile. For example, the grating can be used for long-working distance coupling into single mode fiber (SMF) with 10.4 μm mode field diameter (MFD) and NA of 0.14, by designing the focal distance to be at ~340 μm. As the LAGC requires a challenging fabrication process, our fabrication strategy will provide a solution for achieving accurate patterns with limited errors in periodicity, gap size, and etch depth. To improve performance, a design process with a feedback loop can be implemented, where the observed variation can be integrated into the design equation.

## 3. Conclusions

The Large Area Grating Coupler (LAGC) allows for efficient out-of-plane focusing and intensity control of near infrared (NIR) light ($\lambda$~1.550 μm) from a large area (~100 μm length scale) into a small focal spot of < 10 μm in size. Our method allows the grating to precisely focus the light into a desired spot without the need of an additional lens. The grating periodicity and duty cycle are apodized, as well as curved to achieve the desired focusing effect and intensity modulation. Due to the large dimensions, precise fabrication of such grating is challenging. We introduce new strategies for the fabrication process so that large-area structures can be patterned with minimal errors. By dividing the grating into different patterning sections, a 95.8 μm-long and 108 μm-wide (on widest groove) grating was fabricated with the grating teeth having smallest gap size of 16 nm with a width of ~90 μm, corresponding to an aspect ratio of ~5600:1. The grating exhibits simulated free-space focusing efficiency of 33% (-4.81 dB) with fabricated device efficiency of 22% (-6.58 dB). Further reduction in intensity loss can be expected by designing reflectors underneath the grating and tuning the BOX layer thickness. Future work will focus on implementation of the design and fabrication technique for high-power applications and coupling into a fiber.

## 4. Methods

*4.1 Design Calculation and Generation*

The design periodicity, curve radius, and duty cycle were calculated using MATLAB (Mathworks, Massachusetts, USA). The GDSII layout was generated based on the calculated design using Raith_GDSII MATLAB toolbox v 1.2 (National Institute for Nanotechnology, Alberta, Canada).

*4.2 Substrate Preparation*

SOI wafer (500 nm device thickness) was diced into 1 cm x 1 cm dies. The die was cleaned via acetone bath, isopropyl alcohol (IPA) bath, deionized (DI) water rinse, and followed by a blow-dry with $N_2$. To dry potentially remaining solvents, the substrate was baked on hot-plate set at 140 º C for 120 s.

*4.3 HSQ Coating and Cross-linking*

4% hydrogen silsesquioxane (HSQ) (Dow Corning XR1541) was spin coated onto Silicon-on-Insulator substrate at 5000 rpm for 60 s with a ramped-up time of 2 s, to achieve a resist film thickness of 50 nm. The sample was heated on a hot-plate to promote the cross-linking process. The temperature was set at 550 ºC for 90 minutes, with an actual temperature of 420 ºC as measured by an infrared (IR) thermometer. The sample was covered by a beaker glass to produce more uniform ambient temperatures. After heating, the sample was immediately put on a stainless-steel plate at room temperature for 1 minute to cool down.

*4.4 Electron Beam Lithography*

Polymethyl methacrylate (PMMA) A4 positive tone resist from MicroChem was spin coated onto the HSQ coated sample at 4000 rpm for 60 s with a ramped-up time of 2 s, to achieve a resist film thickness of 200 nm. To increase resistivity, PMMA was pre-baked at 180º C for 100 s (hot plate set at 200 ºC). HSQ and PMMA thicknesses was measured using the Filmetrics F20 thin-film measurement system.

We used the Raith eLine Plus EBL system (Raith GmbH) to pattern the layout with an accelerating voltage of 30 kV, a field size of 200 μm x 200 μm, and a step size of 5 nm. The structures were patterned through multiple exposure methods. The high-resolution section was patterned via exposure of single pixel lines with dose a of 200 pC/cm. To achieve larger gaps, structures were designed through combination of single pixel lines, with boundary dose of 400 pC/cm and inner dose of 150 pC/cm. To reduce patterning time, the end section of the gratings with the largest gaps was patterned using area pattering with a dose of 200 μC/cm$^2$. We developed the structures by sonication in a solution of MIBK:IPA (1:3) at room temperature for 70 s. The sample was then immediately blow dried with a steady stream of $N_2$ gas.

*4.5 Pattern Transfer*

**Table 1. Summary of Etching Process**

|  | **HSQ** | **Si** |
| --- | --- | --- |
| **Gas** | CF4 | HBr |
| **Selectivity** | HSQ/PMMA= 1/2.5 | Si/HSQ= 10.5/1 |
| **Etch rate** | 1.1 nm/s | 1.22 nm/s for waveguide<br>0.85 to 1.06 nm/s for grating |

Pattern transfer was done using Inductive Coupled Plasma-Reactive Ion Etcher (ICP-RIE) (Apex SLR ICP, Advance Vacuum Systems, Lomma, Sweden). Note that we have already optimized the ICP-RIE etching parameters, both for oxide resist and Si, to improve etch rate, etch profile, and minimize etch depth variation. This is including tuning of gas flow, pressure, ICP Power, and Bias Power [34–38]. The pattern was transferred from PMMA mask into cross-linked HSQ via CF$_4$ dry-etch, with 45 sccm gas flow. The chamber pressure was set at 5 mTorr, with 300 W ICP Power and 30 W Bias Power. The substrate table temperature was set to 20º C. We observed an HSQ etch rate of 1.1 nm/s with HSQ/PMMA selectivity= 1/ 2.5, thus an exposure time of 46 s will allow the 50 nm HSQ to be fully etched with remaining PMMA layer of 75 nm thickness.

The pattern was further transferred to SOI Si device layer via HBr dry etch, with 100 sccm gas flow. The chamber pressure was set at 10 mTorr, with 200 W ICP power and 100 W bias power. The substrate table temperature was set to 10º C. For the waveguide region, we observed a Si etch rate of 1.22 nm/s with Si/HSQ selectivity= 10.5/1. To fully etch a 500 nm thick Si layer, the etch time was set to 410 s. The grating region, with the gap size varying from 16 to 221 nm, was etched for 100 s, exhibiting an etch rate of 0.85 to 1.06 nm/s. The fabricated structures were imaged with a JSM-7600F field-emission SEM (Jeol, Tokyo, Japan) at 5 or 10 kV with a working distance of 5 mm. The etched depths were measured using the Atomic Force Microscopy (AFM) (MFP-3D Origin, Asylum Research) and verified with cross-sectional SEM images of test grating structures that had longer grating width to facilitate cleaving.

The remaining oxide mask after the first etch (grating partial etch) was removed using the buffered oxide etch (BOE) process, by dipping the sample into buffered hydrogen fluoride (HF) for 60 s. The sample was then rinsed thoroughly with DI water. To test that there is no oxide mask remaining, we analyzed the hydrophobic properties of the sample, as the surface should appear hydrophobic if oxide layer is completely removed. This oxide removal process was not needed after the second etch (waveguide layer full etch) as the remaining oxide was only ~4 nm in thickness, close to the native oxide thickness, and would also create an undercut structure on the BOX layer if performed.

*4.6 2-D FDTD Simulation*

A commercial finite-difference time domain (FDTD) solver (Lumerical Solutions Inc., Vancouver, Canada) was used to model the focusing effect. The simulated structure was the full grating with 201 grooves, on a SOI substrate with a 500 nm thick Si device layer and a 3 μm thick BOX layer. The simulation span was set to 122 μm in the x direction (propagation direction) and 120 μm in the y direction. The refractive indices for Si and SiO$_2$ were selected from Palik (3.476 and 1.444 respectively) [39] from the Lumerical FDTD Material database, while the background index was set to 1. An automatically generated non-uniform mesh, with a minimum size of 10 nm, was used for the simulation, allowing the grooves to be resolved, while shortening the simulation time. Perfectly matched layer (PML) boundaries were used for all of the boundaries. The source was set as a mode source (TE polarized) normal to grating propagation direction and centered at 1.550 um with a 0.1 μm bandwidth. Field monitors were placed above and below the Si device layer, before and after the grating, and on the far-field (as well as at the focal plane) to record the electric field and power distributions. The focusing efficiency was calculated by comparing the power at the focal point with the power of the guided mode in the input waveguide.

**Funding**

Digital Manufacturing and Design Centre (DManD) (RGDM1830303).

**Acknowledgement**

We would like to thank Singapore University of Technology and Design (SUTD), SUTD-MIT International Design Centre (IDC), and Institute of Material Research and Engineering (IMRE) for providing tools and facilities.

## Disclosure

The authors declare no conflicts of interest.